\documentstyle[sprocl]{article}

\def\PRD{{\it Phys. Rev.} D}

\def\be{\begin{equation}}
\def\ee{\end{equation}}
\def\bea{\begin{eqnarray}}
\def\eea{\end{eqnarray}}

\begin{document}

\title{LORENTZ VIOLATION AND HAWKING RADIATION}

\author{T. JACOBSON}

\address{Department of Physics,
University of Maryland, College Park\\ 
MD 20742-4111, USA\\E-mail: jacobson@physics.umd.edu}


\maketitle

\abstracts{Since the event horizon of
a black hole is a surface of infinite redshift, it might be
thought that Hawking radiation would be highly sensitive to
Lorentz violation at high energies.  In fact, the opposite is
true for subluminal dispersion.  For superluminal dispersion,
however, the outgoing black hole modes emanate from the
singularity in a state determined by unknown quantum gravity
processes.}


As evidenced by this meeting, there is a growing feeling that
Lorentz invariance should be questioned and tested.  Tentative
results from from quantum gravity and string theory hint that the
ground state may not be Lorentz invariant, and new techniques are
allowing for experiments with unprecedented precision and
astrophysical observations at unprecedented high energies. 
Moreover, due to the unboundedness of the boost parameter, {\it
exact} Lorentz invariance, while mathematically elegant, is
unverifiable and therefore suspect.

Most work exploring possible Lorentz violation has focused on
non gravitational physics---i.e. flat spacetime.  Much less has
been done to investigate Lorentz breaking in curved spacetime, or
in gravitational phenomena themselves.  The event horizon of a
black hole, being a surface of infinite redshift, is a particularly
good probe of the limits of boost invariance at high energies. 
Since Hawking radiation emerges from the vicinity of the horizon,
it is therefore interesting to investigate the impact of Lorentz
violation on Hawking radiation.  To pursue this question,
however, one must first address another question: {\it if Lorentz
violation exists, what becomes of general relativity?}

In flat spacetime, Lorentz breaking is described by couplings to
constant symmetry breaking tensors $V_a$, $W_{ab}$, ...  . To
formulate Lorentz breaking in a curved spacetime without
destroying general covariance such tensors must become {\it
dynamical tensor fields} fields that satisfy (effective) field
equations.  An implementation of this, with a unit timelike
vector $u^a(x)$ as the Lorentz breaking field, is the subject of
another contribution to these proceedings \cite{MJ,JM}.  Such a
field preserves rotation invariance and, since it is a {\it unit}
vector, contains no information other than the determination of a
preferred rest frame at each point of spacetime.  For the present
purposes let us just imagine that some such implementation
exists.

Hawking radiation is thermal radiation emitted from a black hole. 
For a black hole with surface gravity $\kappa$, the Hawking
temperature is $T_H=\hbar\kappa/2\pi c$.  A nonrotating black
hole has a surface gravity $c^ 2/2R_S$, where $R_S$ is the
Schwarzschild radius, hence the thermal wavelength is
$\lambda_H=8\pi^2 R_S$.  This is the typical wavelength of the
Hawking radiation far from the black hole, after it has climbed
out of the potential well of the hole.  Following the radiation
backwards all the way to the horizon, however, it recedes into
the vacuum fluctuations and the wavelength goes to zero as
measured by an observer falling freely across the event horizon. 
The Hawking radiation therefore originates from Planck (or
trans-Planck) scale vacuum fluctuations, which may be governed by
completely unknown physics if Lorentz invariance is broken.

The essence of the Hawking effect is independent of field mass or
interactions, hence it is adequate for the present purposes to
consider a free massless field.  The propagation of the field
near the horizon can be inferred in the WKB approximation from
the dispersion relation, and Lorentz violation can be
parameterized by modifications of the dispersion relation. 
Consider for example dispersion relations of the form
\begin{equation}
\omega=k +\xi k^{1+n}/k_P^{n},
\label{dr}
\end{equation}
where $\omega$ is the frequency, $k$ is the magnitude of the wave
3-vector, $k_P$ is the Planck wave vector, and units with $c=1$
are employed.  The dimensionless parameter $\xi$ controls the
amount of Lorentz violation and the integer $n$ determines its
$k$ dependence.  The case $n=0$ is scale independent and
corresponds to the renormalizable Lorentz violation one has in
the standard model extension.\cite{Kos} In the case $n>0$ the
Lorentz violation grows stronger at higher energies.  The group
velocity $v_g=d\omega/dk$ for the dispersion relation (\ref{dr})
is given by
\begin{equation}
v_{g}= 1 + 
\xi(1+n)(k/k_P)^{n},
\label{vg}
\end{equation}
which is superluminal if $\xi>0$ and subluminal if $\xi<0$.  The
frequency and spatial wave 3-vector in (\ref{dr}) are defined
with respect to the preferred frame $u^a$.  Explicitly,
$\omega=u^ak^{(4)}_a$ and $k^{(3)}_a= (\delta_a{}^b-u_a
u^b)k^{(4)}_b$.  The dispersion relation is imported into curved
spacetime by using the vector field $u^a(x)$. 

To investigate the consequences of Lorentz breaking for the
Hawking effect it is first of all necessary to specify the
profile of the preferred frame in the black hole spacetime.  In
spherical symmetry a natural frame is the radial free-fall frame
which is determined by the geodesics that are asymptotically at
rest at spatial infinity and fall freely across the horizon. 
Whether this particular frame would be dynamically selected by
the effective field theory is unknown.  However it seems
reasonable to assume, as I will here, that the preferred frame is
well-behaved at the horizon.  This means that it has finite
velocity with respect to the free fall frame and that the
invariant 4-acceleration is not large compared to the surface
gravity $\kappa$.

We can now examine how Lorentz violation might affect the
propagation of fields near a black hole horizon.  Consider first
the $n=0$ case, for which the group velocity is the
$k$-independent constant $1+\xi$.  For negative $\xi$ this is
{\it less} than the speed of light, so the effective horizon
moves out.  In fact the dispersion relation is identical to that
of a Lorentz invariant massless field coupled to the metric
$g'_{ab}=g_{ab}+(2\xi+\xi^2) u_au_b$.  The light cones of
$g'_{ab}$ are `narrower' than those of $g_{ab}$ so the horizon is
moved out relative to that of $g_{ab}$.  A matter field with such
dispersion would exhibit the usual Hawking effect, with a
temperature determined by the surface gravity of the $g'_{ab}$
horizon.  For positive $\xi$ the group velocity is instead {\it
greater} than the speed of light, so the $g'$ light cones are
`opened up' and the horizon moves {\it inward}.

For $n>0$ the situation is much more interesting.\cite{River} Let
us first consider the case where $\xi$ is positive.  Then the
group velocity is superluminal (as with $n=0$), and grows with
$k$.  This means that as an outgoing wavepacket is followed
backwards in time toward the horizon, it speeds up and crosses
the horizon, continuing to go faster, until it approaches the
curvature singularity inside the black hole.  In other words, the
outgoing mode emerged from the vicinity of the singularity and
propagated superluminally out across the horizon, finally
redshifting enough to slow down to the speed of light.

Needless to say, this is very different from the usual Hawking
effect.  In the usual case, the outgoing mode is stuck just
outside the horizon, exponentially blueshifting backwards in
time.  The Hawking effect is deduced from a vacuum condition
imposed on the quantum field near the horizon, namely that
outgoing modes with frequency $\omega\gg\kappa$ are in their
ground state as seen by a free-fall observer.  This physically
reasonable condition need not refer to any Planck or trans-Planck
scale quantities.  It merely expresses the hypothesis that after
the black hole formed, there is no process going on that could
excite these modes.  (However, due to the continued blueshift
backwards in time, it cannot be {\it deduced} from the initial
conditions before the collapse that formed the black hole without
reference to trans-Planckian quantities.)  In this sense the usual
Hawking effect is a robust prediction of Lorentz invariant field
theory.

In the presence of $n>0$ superluminal dispersion, however, the
outgoing modes emerge from the vicinity of the singularity inside
the black hole where unknown physics takes place.  If they emerge
in their ground state then the usual Hawking effect would occur. 
But something entirely different might happen, and indeed seems
more plausible, since the modes propagate through a region of
diverging spacetime curvature.

Finally, let us consider the case of negative $\xi$.  The group
velocity is then subluminal, and decreases with increasing $k$. 
In this case, as first shown by Unruh \cite{Unruh95}, an outgoing
wavepacket originates as an ingoing wavepacket, containing
typical wavevectors of order $k_P$, which is outgoing with
respect to the preferred frame but not fast enough to compensate
for the infalling of that frame.  Thus the wavepacket is dragged
towards the horizon.  As this happens the wavepacket redshifts,
and near the horizon it finally redshifts enough for its group
velocity to exceed the infall speed of the preferred frame.  At
this point it turns around and finally climbs away from the
horizon, continuing to redshift on the way out.  This sort of
continuous evolution of a mode from one type to another, with a
change of group velocity, in response to propagation through an
inhomogeneous medium, is well known is various other areas of
physics.  It is called ``mode conversion" in the plasma physics
literature.

Now what does this mode conversion imply for the Hawking effect? 
Actually, it makes no difference!  The same physical reasoning as
in the Lorentz invariant case supports the hypothesis that the
outgoing modes with $\omega\gg\kappa$ near the horizon are in
their ground state as seen by free fall observers.  The only
difference is in the {\it ancestry} of the modes.  In the Lorentz
invariant case the modes have trans-Planckian ancestry due to the
continued exponential blueshifting at the horizon, and they
originate as ingoing modes that arrive at the horizon just before
it forms from the collapse that produced the black hole.  In the
dispersive case on the other hand, the ancestor is only
Planckian, and it originates as an ingoing (or rather, dragged-in
outgoing) mode that arrives at the horizon {\it after} the black
hole forms.  It was shown numerically in Ref.  \cite{Unruh95}
(using a dispersion relation for which the group velocity
vanishes at large $k$) that if these ingoing modes are in their
ground state far from the black hole then the usual Hawking flux
emerges as long as $\kappa\ll k_P$.

A weakness in the above account is that the model breaks down if
one attempts to trace the ingoing mode all the way back out to
spatial infinity, since the wavelength gets arbitrarily
blueshifted in the process.  An improved model would start from
some sort of discreteness for spacetime with a physically
sensible short distance cutoff.  The Hawking effect has been
studied in such a model, in which space is treated as a lattice
that falls freely into a black hole in 1+1
dimensions.\cite{JMlattice} 

Linear fields propagating on the lattice naturally have
subluminal dispersion at high wavevectors.  For a lattice with
spacing $a$ the dispersion relation is $\omega
=(2/a)|\sin(ka/2)|$.  Wavevectors differing by $2\pi/a$ are
identified, so the independent modes are labelled by the
so-called Brillouin zone $(-\pi/a,\pi/a)$.  The group velocity
vanishes for $k=\pi/a$, and it is negative for yet higher
wavevectors, which are identified with negative wavevectors.  The
wavepacket propagation picture discussed above suggests that mode
conversion would take place outside a black hole horizon, with an
outgoing wavepacket of small positive $k$ components arising from
an ingoing packet of $k$ components greater than $\pi/a$.  (This
is analogous to a Bloch oscillation, which occurs when a charged
particle in a periodic potential is subjected to a uniform
electric field.)  This behavior was confirmed by numerical
propagation of a scalar field wavepacket on a falling lattice
using the lattice field equation.  Moreover, assuming the ingoing
modes are in their quantum ground states, the Hawking radiation
thermal occupation numbers for the outgoing modes were found as
long as the lattice spacing was small compared with the length
scale $\kappa^{-1}$ of the black hole.

In this falling lattice model the slow time-dependent spreading
of the lattice points is critical to producing the net frequency
change of the mode that we know must occur if the Hawking effect
is to take place.  Such rarefaction of the lattice is an
artificial feature of this simple 1+1 dimensional model however. 
Moreover, in any process involving modes with Planckian
wavevectors, the linear model of the test field is surely
unjustified.  The back-reaction on the quantum gravity vacuum
must be essential.  It seems unavoidable to suppose that the
fluctuations of the actual quantum gravity vacuum obviate the
need for the explicit time dependence of the falling lattice, and
that the dynamics of this vacuum provides a non-linear way of
partially converting ingoing fluctuations to outgoing ones.

In conclusion, Lorentz violation is entirely compatible with the
usual understanding of black holes in general relativity,
including the Hawking effect, except in the case of superluminal
dispersion with a group velocity that grows without bound at high
wavevectors.  In that case, modes emanating from the vicinity of
the singularity would emerge from the black hole, their state
having been determined by unknown quantum gravity processes.  \\

\noindent {\it This work was supported in part by the 
NSF under grant No.  PHY98-00967.}

\end{document}